\documentclass[aps,prl,twocolumn,superscriptaddress,showpacs]{revtex4}

\usepackage{graphicx}
\usepackage{amsmath}
\usepackage{latexsym}

\begin{document}

\title{Anisotropic de Gennes narrowing in confined fluids}

\author{Kim~Nyg{\aa}rd} 
\email[Corresponding author: ]{kim.nygard@chem.gu.se}
\affiliation{Department of Chemistry and Molecular Biology, University of Gothenburg, 
SE-41296 Gothenburg, Sweden}

\author{Johan~Buitenhuis}
\affiliation{Forschungszentrum J\"{u}lich, ICS-3, D-52425 J\"{u}lich, Germany}

\author{Matias~Kagias}
\affiliation{Paul Scherrer Institut, CH-5232 Villigen PSI, Switzerland}
\affiliation{Institute for Biomedical Engineering, UZH/ETH Z{\"u}rich, CH-8092 Z{\"u}rich, Switzerland}

\author{Konstantins~Jefimovs}
\affiliation{Paul Scherrer Institut, CH-5232 Villigen PSI, Switzerland}
\affiliation{Institute for Biomedical Engineering, UZH/ETH Z{\"u}rich, CH-8092 Z{\"u}rich, Switzerland}

\author{Federico~Zontone}
\author{Yuriy~Chushkin} 
\affiliation{European Synchrotron Radiation Facility, 71 Avenue des Martyrs, F-38000 Grenoble, France}

\date{\today}

\begin{abstract}
The collective diffusion of dense fluids in spatial confinement was studied by combining high-energy 
(21~keV) x-ray photon correlation spectroscopy and small-angle x-ray scattering from colloid-filled 
microfluidic channels. We found the structural relaxation in confinement to be slower compared to bulk. 
The collective dynamics is wave vector dependent, akin to de Gennes narrowing typically observed 
in bulk fluids. However, in stark contrast to bulk, the structure factor and de Gennes narrowing in 
confinement are anisotropic. These experimental observations are essential in order to develop a 
microscopic theoretical description of collective diffusion of dense fluids in confined geometries.
\end{abstract}



\maketitle


Dense fluids in narrow spatial confinement exhibit complex microscopic ordering, due to competing 
packing constraints imposed by the confining surfaces and the other particles in the 
system~\cite{henderson}. As a result confinement alters the fluids' dynamic properties~\cite{granick91}, 
and is thus highly relevant in a wide range of scientific phenomena and technological applications, 
including water transport in molecular sieves, hindered motion of colloidal dispersions in porous matrices, 
the glass transition in thin films, lubrication, and various micro- and nanofluidic applications. 
State-of-the-art experiments~\cite{nugent07} and simulations~\cite{mittal08} have shown that 
packing constraints in confined geometries lead to position-dependent diffusion of dense fluids, 
thereby highlighting the close connection between microscopic structure and dynamics. 
Correlations between structural quantities such as the excess entropy, on the one hand, and 
dynamic properties such as single-particle diffusion~\cite{mittal06} or structural 
relaxation~\cite{ingebrigtsen13}, on the other hand, have also been reported. However, nanoscopically 
confined dense fluids exhibit complex dynamical behavior~\cite{lang10}, and a conceptually simple 
mechanistic picture in terms of the microscopic structure is still missing. 

An established description of microscopic dynamics in bulk is provided via a phenomenon known as 
{\em de~Gennes narrowing}~\cite{degennes59}, according to which the wave vector dependent 
collective diffusion coefficient $D({\bf q})$ scales as the inverse of the structure factor $S({\bf q})$. 
In essence, density fluctuations with a wave vector ${\bf q}$ corresponding to a maximum in the 
structure factor have a low free energy cost and thus decay slowly, providing a straightforward 
physical connection between microscopic structure and dynamics. This scheme has been used to 
analyze collective diffusion in bulk systems ranging from colloidal dispersions~\cite{banchio06} to 
glass-forming silicates~\cite{ruta14}, and has even been successfully applied to the relative motion 
of protein domains~\cite{hong14}. However, its validity in confinement is yet to be demonstrated, whether 
experimentally, theoretically, or by simulations. Spatial confinement induces anisotropy in the fluid's pair 
correlations~\cite{kjellander91} and the ensuing structure factor~\cite{nygard13}. If de~Gennes narrowing 
holds in confinement, then one would observe anisotropic wave vector dependent collective diffusion. 

In this Letter, we study the connection between the microscopic structure and dynamics of dense fluids 
confined between planar walls at close separation, i.e., in a narrow slit geometry. We have recently 
developed a unique scheme to experimentally determine anisotropic structure factors of confined 
fluids, based on small-angle x-ray scattering (SAXS) from colloid-filled micro- or nanofluidic 
channel arrays~\cite{nygard09}, in semi-quantitative agreement with {\em ab initio} theoretical 
predictions~\cite{nygard12,nygard16}. 
Here we extend the methodology for simultaneous determination of the 
static structure factor and the wave vector dependent collective diffusion coefficient of a charge-stabilized 
colloidal dispersion in narrow confinement, by carrying out high-energy SAXS and x-ray photon 
correlation spectroscopy (XPCS) experiments. We provide the first observation of anisotropic 
de~Gennes narrowing in confined fluids, with the structural relaxation being 
highly anisotropic and significantly slower compared 
to bulk. These experimental findings, which establish a connection between structure and dynamics 
at the fundamental level of pair densities, 
are crucial for the development of a microscopic theory of dynamics in dense confined fluids.


Following Refs.~\cite{nygard09,nygard12,nygard16}, 
we used a colloidal dispersion confined in a 
specifically designed microfluidic channel array as a model for confined fluids. The colloid consisted 
of charge-stabilized spherical silica particles dispersed in ethylene glycol, with a particle volume fraction of 
$\phi = 0.168$ in bulk. The average diameter of the particles was $\sigma = 182$~nm and the 
polydispersity $\Delta\sigma / \sigma  = 1.5\%$, as determined by SAXS from a dilute bulk dispersion. 
Details on the synthesis of the colloidal dispersion can be found elsewhere~\cite{gapinski09}.  

The microfluidic container, in turn, consisted of a periodic array of one-dimensional channels. 
The channel array was made into a $300~\mu$m thick silicon wafer 
by electron-beam lithography and KOH etching following Ref.~\cite{mao09}. 
It had a period of $2~\mu$m, a depth of $\approx 18~\mu$m, and a channel width of $H = 490$~nm 
(i.e, $H \approx 2.7\sigma$), as determined by scanning electron microscopy. The resulting confining 
walls were structureless on the length scales relevant for the colloidal dispersion, facilitating studies on the 
connection between microscopic structure and dynamics. We also prepared space for a bulk fluid 
reservoir, allowing us to collect bulk data from the same sample cell. Finally,  we covered the channel 
array by a $500~\mu$m thick glass plate in order to prevent evaporation of the solvent and to facilitate 
attachment of syringes for filling the fluid, resulting in a $3~\mu$m thin fluid film between the channel 
array and the glass cover.  

We carried out the combined XPCS and SAXS experiment at the beamline ID10 of the European 
Synchrotron Radiation Facility (ESRF). Our key technical advance is the use of high-energy 
incident x-rays, thereby minimizing absorption in the beam path and allowing the XPCS experiment 
on the colloidal dispersion sandwiched between the silicon wafer and the glass cover. The incident x-ray 
beam had an energy of $\hbar\omega = 21$~keV, a size of $10 \times 10~\mu$m$^2$ at the sample 
position, and impinged parallel to the confining walls. To maximize the intensity the x-ray beam was 
focused onto the sample plane and an evacuated flight tube was placed between the sample and the 
detector to minimize the background from air scattering. Scattered x-rays were collected in transmission 
mode 5.3~m behind the sample in a twofold manner as follows. First, we obtained an overview of the 
static structure using the two-dimensional single-photon-counting CdTe MAXIPIX detector 
($256 \times 768$ pixels with a size of 
$55 \times 55~\mu$m$^2$ each)~\cite{ponchut11}. Then, we collected static SAXS and dynamic XPCS 
data simultaneously, in selected scattering vector directions with respect to the confining channels, using a 
$0.1 \times 0.1$~mm$^2$ slits opening in front of a point detector, a Cyberstar scintillation counter 
connected to a Flex hardware correlator. The experiment was carried out at room temperature.

\begin{figure}
\centering\includegraphics[width=7.8cm]{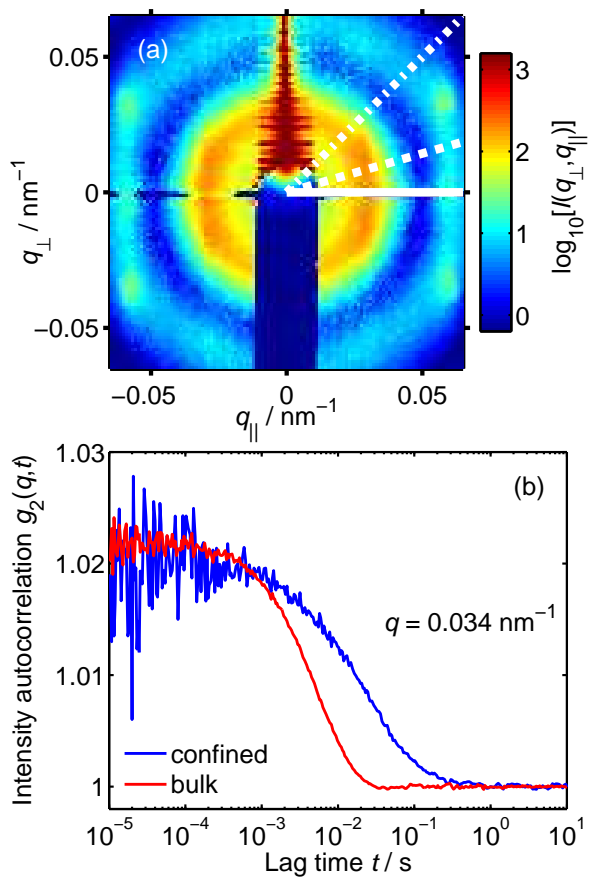}
\caption{(color online) 
(a) Static SAXS intensity $I({\bf q})$ obtained from the confined fluid, presented 
on a log scale as a function of 
scattering vector components parallel ($q_{\parallel}$) and perpendicular ($q_\perp$) to the confining 
channels. The solid, dashed, and dashed-dotted white lines depict the parallel ($0^\circ$), non-parallel 
($16^\circ$), and diagonal ($45^\circ$) directions (wrt. the confining channels) in which we collect 
dynamic data. The dark red and blue features at $q_\parallel = 0$ are due to diffraction from the 
confining container and the central beam stop, respectively, and should be neglected in the 
discussion~\cite{note_diff}. 
(b) Intensity autocorrelation functions $g_2(q,t)$ collected from the confined (blue) 
and bulk (red) colloidal dispersion at a scattering vector magnitude of $q= 0.034$~nm$^{-1}$. The 
data from the confined fluid are collected at an angle of $16^\circ$ with respect to the confining channels 
[denoted non-parallel; 
see dashed line of panel~(a) 
for a graphical definition]. 
\label{fig:saxs_cf}
}
\end{figure}

First, we discuss the static SAXS intensity $I({\bf q})$ obtained from the confined fluid, shown in 
Fig.~\ref{fig:saxs_cf}(a). 
These data, like all data presented in this Letter, have been verified in several repeat measurements. 
The data are proportional to the anisotropic structure 
factor~\cite{note_SAXS}, 
$S({\bf q}) = 1 + \frac{1}{N} \int \int n({\bf r})n({\bf r}')h({\bf r},{\bf r}') 
e^{i {\bf q} \cdot ({\bf r} - {\bf r}')} d{\bf r} d{\bf r}'$,  
where ${\bf q}$ denotes the wave/scattering vector, $N$ the total number of particles, $n({\bf r})$ the 
number density profile across the confining slit, $h({\bf r},{\bf r}')$ the pair correlation function, and the 
integration is carried out over particle positions ${\bf r}$ and ${\bf r}'$. For the present case of a fluid in 
slit confinement, $S({\bf q})$ is essentially given by the Fourier transform of the pair density correlation 
$n({\bf r})h({\bf r},{\bf r}')$, ensemble averaged over particle positions across the confining channel
(for illustrative real-space representations, see Ref.~\cite{nygard13}). 
In stark contrast to isotropic bulk fluids, the SAXS intensity obtained from 
the confined fluid is strongly anisotropic, directly demonstrating the strong anisotropy in the structure 
factor and the underlying pair correlations.

Structural relaxation in the fluid was probed by XPCS which measures the 
intensity autocorrelation function
\begin{equation}
g_2({\bf q}, t) = \frac{\langle I({\bf q}, t_0) I({\bf q}, t_0+t) \rangle_{t_0}}
{\langle I({\bf q}, t_0) \rangle_{t_0} \langle I({\bf q}, t_0+t) \rangle_{t_0}}, 
\end{equation}
where $\langle \cdots \rangle_{t_0}$ denotes a temporal average over all times $t_0$~\cite{grubel}. 
The $g_2({\bf q}, t)$ curves obtained from both the confined and the bulk fluid at a scattering vector 
magnitude of $q= 0.034$~nm$^{-1}$ are presented in Fig.~\ref{fig:saxs_cf}(b), 
the former in the non-parallel direction as defined in panel~(a). 
These correlation functions differ in two ways. First, the dynamics is significantly 
slower in confinement compared to bulk at this particular scattering vector. Second, the 
shape of the correlation function in confinement is stretched, in contrast to bulk. 

The intensity autocorrelation functions of 
Fig.~\ref{fig:saxs_cf}(b) 
are related to the fluids' density fluctuations via the Siegert relation, 
$g_2({\bf q}, t) = 1+\gamma \vert f({\bf q}, t)\vert^2$, 
where $\gamma$ denotes the experimental contrast and $f({\bf q},t)$ the normalized intermediate 
scattering function~\cite{grubel}. For quantitative analysis we fit the correlation functions with 
the Siegert relation 
using the phenomenological Kohlrausch--Williams--Watts 
(KWW) 
exponential expression, 
$f({\bf q},t) = \exp \left\{-[t/\tau({\bf q})]^{\beta}  \right\}$, 
which is commonly applied to describe dynamics in disordered systems. Here $\tau ({\bf q})$ denotes 
the wave vector dependent relaxation time and $0 < \beta \leq 2$ the shape factor, with $\beta < 1$  
corresponding to the stretched correlation function. From the fits of the $g_2({\bf q},t)$ curves, 
we obtain nearly constant $\beta \approx 0.95$ for the bulk, 
$\approx 0.88$ for the fluid film, as well as   
$\approx 0.72$, $\approx 0.70$, and $\approx 0.78$  
for the confined fluid in diagonal, non-parallel, and parallel directions 
[see Fig.~\ref{fig:saxs_cf}(a) for definitions].    
To account for the stretching we consider throughout this study an average relaxation time, 
$\langle \tau ({\bf q}) \rangle \equiv \int_0^\infty f({\bf q},t) dt = \tau ({\bf q}) \beta^{-1}\Gamma^* (\beta^{-1})$, 
where $\Gamma^*$ denotes the Gamma function. 

Stretching of the correlation function indicates a distribution of relaxation times coming either from 
the sample itself or as a result of the experimental conditions. Because our sample contains 
monodisperse particles, we attribute the weak stretching in bulk ($\beta \approx 0.95$) to the averaging 
of relaxation times over a large slit opening (0.1$\times$0.1 mm$^2$) in front of the point detector. The 
data from the confined fluid ($18~\mu$m thick) also contain a contribution from the 3 $\mu$m thick fluid film 
on top of the channel array~\cite{note_reservoir}, which could 
explain the stronger stretching ($\beta \approx 0.75$) compared to bulk. 
However, the enhanced stretching observed in the film ($\beta \approx 0.88$) indicates that  confinement 
indeed induces stretched dynamics in our fluid. 
Our present experimental design does not allow XPCS experiments on the confined fluid alone, and in 
the future a new cell geometry will be envisaged to study this effect. 
Importantly, the contribution from the fluid film does not depend on the azimuthal angle, and cannot 
therefore explain the main observation in our study -- anisotropic de Gennes narrowing.

\begin{figure}
\centering\includegraphics[width=7.8cm]{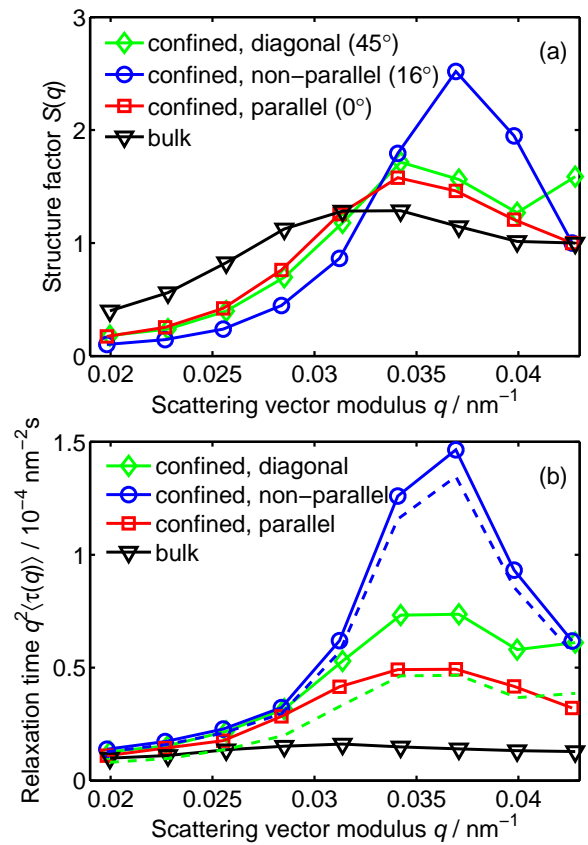}%
\caption{(color online) 
(a)~Static structure factor $S({\bf q})$ obtained from the static SAXS data. 
(b)~Reduced wave vector dependent relaxation 
time $q^2 \langle \tau ({\bf q}) \rangle$, as determined from the dynamic XPCS data using the 
KWW model (see text for details). Data are shown for the confined fluid in diagonal (green diamonds),  
non-parallel (blue circles), and parallel (red squares)  directions as well as for the bulk fluid (black triangles). 
The dashed lines correspond to reduced relaxation times for the confined fluid, normalized to 
the effective viscosity of the solvent in the parallel direction (see text for details).  
\label{fig:xpcs} 
} 
\end{figure}

Having introduced the static structure factor $S({\bf q})$ and the wave vector dependent relaxation 
time $\tau({\bf q})$, we can now address the main question of the present study -- how does the 
microscopic structure of the confined fluid affect its collective diffusion? For this purpose we present in 
Fig.~\ref{fig:xpcs}(a) the structure factor $S({\bf q})$, as obtained by dividing the SAXS intensity 
$I({\bf q})$ by the form factor of a dilute dispersion. The data are shown for four 
different cases: the confined fluid in diagonal, non-parallel, and parallel directions 
as well as the bulk fluid. We note two interesting features in the data. First, the peak 
positions and intensities differ between bulk and confinement, 
implying a denser packing of particles in the latter case. 
A possible explanation may be the formation of dense particle monolayers near each confining wall, 
which has been observed previously for charge-stabilized colloids in confinement~\cite{satapathy08}. 
Second, the data for the confined fluid are direction dependent, as inferred already from the anisotropic 
SAXS data in Fig.~\ref{fig:saxs_cf}(a), 
demonstrating an anisotropic $S({\bf q})$ and implying ensuing anisotropic properties of the confined fluid. 
Such anisotropy in $S({\bf q})$ has previously been observed for both charged~\cite{nygard09} and 
hard-sphere colloids~\cite{nygard12,nygard16} in spatial confinement.

The wave vector dependent decay rate of density fluctuations, $\Gamma ({\bf q}) = 1 / \tau({\bf q})$, 
can be cast in the form of the collective diffusion coefficient, $\Gamma ({\bf q}) = q^2D({\bf q})$. On 
the other hand, de~Gennes narrowing implies that the latter behaves as the inverse of the structure 
factor, $D({\bf q}) \propto 1/S({\bf q})$. We thus expect the reduced wave vector dependent relaxation time to 
be proportional to the structure factor, $q^2 \tau ({\bf q}) \propto S({\bf q})$~\cite{grubel}. This behavior 
is indeed observed in Fig.~\ref{fig:xpcs}(b), where we present 
$q^2 \langle \tau ({\bf q}) \rangle$ for the same four 
systems as in Fig.~\ref{fig:xpcs}(a); both for bulk and confinement we observe a maximum in the 
average reduced relaxation time $q^2 \langle \tau ({\bf q})\rangle$ 
coinciding with the maximum in the structure factor $S({\bf q})$. 
This contrasts with the constant reduced Stokes--Einstein relaxation time 
observed for dilute fluids, 
$q^2 \tau  = 6\pi\eta r_\mathrm{H} / k_\mathrm{B}T$ with $k_\mathrm{B}$ Boltzmann's constant, $T$ 
the absolute temperature, $r_\mathrm{H}$ the particles' hydrodynamic radius ($\approx 91$~nm in 
this study), and $\eta$ the solvent's viscosity~\cite{grubel}. Most importantly, 
the anisotropic $S({\bf q})$ for the confined fluid leads to an anisotropic 
$q^2 \langle \tau ({\bf q})\rangle$, 
with both functions exhibiting a similar behavior as a function of wave vector direction. 
We emphasize that the slower dynamics in the non-parallel compared to the diagonal direction can 
only be explained by anisotropic caging effects due to neighboring particles.  
To the best of our knowledge, this is the first observation of anisotropic de~Gennes narrowing in 
confined fluids, 
highlighting the importance of describing dense confined fluids at the level of anisotropic pair densities. 

The agreement between $q^2 \langle \tau({\bf q})\rangle$ and $S({\bf q})$ in Fig.~\ref{fig:xpcs} is 
only qualitative. 
This is notable when comparing data collected in parallel and diagonal directions; the 
static structure factors are similar, but the dynamics is slower in the latter case. 
A possible explanation is that colloidal dispersions also exhibit hydrodynamic interactions mediated 
by the solvent. In particular, we expect a viscous drag effect due to the presence of a solid surface, 
which depends on both the direction and the particle's distance from the wall (see, e.g., 
Ref.~\cite{holmqvist07}). To estimate the magnitude of this effect, 
we have determined effective $q$-independent viscosities in diagonal, non-parallel, and parallel 
directions following Ref.~\cite{holmqvist07}, assuming a uniform distribution of particle distances from 
the walls and employing the superposition approximation of Ref.~\cite{lin00}. In Fig.~\ref{fig:xpcs}(b) 
we show as dashed lines the reduced relaxation times in diagonal and non-parallel directions, normalized 
to the lower effective viscosity parallel to the confining walls. Although this normalization procedure is only 
approximate, it is clear that the quantitative differences in $q^2 \langle \tau ({\bf q}) \rangle$ between 
diagonal and parallel directions can (at least partly) be attributed to the viscous drag effect. 
Nevertheless, the strongest slowing down in the non-parallel ($16^\circ$) as compared to diagonal  
($45^\circ$) and parallel ($0^\circ$) direction demonstrates that a higher structure factor peak seems 
the strongest predictor for slower dynamics, i.e., de Gennes narrowing. 

In principle our experimental results could be analyzed using mode-coupling theory, which is a dynamic 
counterpart of the theoretical approach used in Refs.~\cite{nygard12,nygard16}. However, although the 
mode-coupling theory has recently been extended to confined hard-sphere fluids, the calculations have
remained difficult and scarce in practice~\cite{lang10,mandal14}. 
Moreover, a quantitative analysis of the data in Fig.~\ref{fig:xpcs}(b) would require the inclusion of both 
hydrodynamic particle--particle~\cite{banchio06} and particle--wall~\cite{holmqvist07} 
interactions, which are highly non-trivial for the present case of a dense fluid in narrow confinement. 
Such an analysis, which may include the counterbalancing of hydrodynamic particle--particle and 
particle--wall interactions~\cite{michailidou09}, has not yet been worked out theoretically for dense 
confined fluids and is beyond the scope of the present study. 
It is our hope that the anisotropic de~Gennes narrowing reported here will both guide theoretical 
development and facilitate interpretation of experiments on structural relaxation in dense confined fluids.  

Finally we comment on a promising future application of the methodology exploited here. Recent 
Monte Carlo simulations and calculations within mode-coupling theory have predicted the existence 
of reentrant glass transitions in confined hard-sphere systems upon varying the surface 
separation $H$~\cite{mandal14}. Our approach provides means to experimentally verify this 
intriguing prediction; by carrying out a combined SAXS/XPCS experiment for dense hard-sphere 
fluids as a function of $H$, one could determine how the onset of the glass transition depends on 
the confining surface separation.

 
In summary, we have simultaneously probed the microscopic structure and wave vector dependent 
collective diffusion of spatially confined fluids, by combining high-energy SAXS and XPCS 
experiments on colloidal dispersion confined in specifically designed microfluidic channel arrays. 
Most importantly, we report the first observation of anisotropic de~Gennes narrowing in confined fluids, 
with the anisotropic structure factor showing up as anisotropic wave vector dependent collective diffusion. 
Our results establish a direct connection between the structure and dynamics in confined fluids 
at the fundamental level of anisotropic pair densities, 
and thereby provide an important conceptual step towards a microscopic description of collective 
diffusion in dense confined fluids.


We thank Marie Ruat for the help with the CdTe MAXIPIX detector and the ESRF for providing 
beamtime. K.N. acknowledges the Swedish Research Council for financial support.


\begin{thebibliography}{28}
\bibitem[{not({\natexlab{a}})}]{note_SAXS}
\bibinfo{note}{Following Refs.~\cite{nygard09,nygard12}, we have carefully
  subtracted scattering data from the fluid film between the channel array and
  the glass cover.}
\bibitem[{not({\natexlab{c}})}]{note_reservoir}
\bibinfo{note}{The scattering volume of the confined fluid is $\approx 50\%$
  larger than that of the fluid film.}
\end{thebibliography}

\end{document}